# Hierarchical nature of hydrogen-based direct reduction of iron oxides

Yan Ma [a,*], Isnaldi R. Souza Filho [a], Yang Bai [a], Johannes Schenk [b], Fabrice Patisson [c], Arik Beck [d], Jeroen A. van Bokhoven [d,g], Marc G. Willinger [e], Kejiang Li [f], Degang Xie [a,h], Dirk Ponge [a], Stefan Zaefferer [a], Baptiste Gault [a,i], Jaber R. Mianroodi [a], Dierk Raabe [a,*]

[a] *Max-Planck-Institut für Eisenforschung GmbH, Max-Planck-Straße 1, 40237 Düsseldorf, Germany*

[b] *Chair of Ferrous Metallurgy, Montanuniversität Leoben, Franz-Josef-Straße 18, 8700 Leoben, Austria*

[c] *Institut Jean Lamour, Labex Damas, Université de Lorraine, 54011 Nancy, France*

[d] *Institue for Chemical and Bioengineering (ICB), ETH Zurich, Vladimir-Prelog-Weg 1-10, 8093 Zurich, Switzerland*

[e] *Scientific Center for Optical and Electron Microscopy (ScopeM), ETH Zurich, Otto-Stern-Weg 3, 8093 Zurich, Switzerland*

[f] *School of Metallurgical and Ecological Engineering, University of Science and Technology Beijing, Beijing 100083, PR China*

[g] *Laboratory for Catalysis and Sustainable Chemistry, Paul Scherrer Institute, 5232 Villigen, Switzerland*

[h] *Center for Advancing Materials Performance from the Nanoscale (CAMP-Nano), State Key Laboratory for Mechanical Behavior of Materials, Xi'an Jiaotong University, Xi'an 710049, PR China*

[i] *Department of Materials, Imperial College, South Kensington, London SW7 2AZ, United Kingdom*

*\* Corresponding authors: d.raabe@mpie.de (Dierk Raabe); y.ma@mpie.de (Yan Ma)*

**Abstract**

Fossil-free ironmaking is indispensable for reducing massive anthropogenic $CO_2$ emissions in the steel industry. Hydrogen-based direct reduction (HyDR) is among the most attractive solutions for green ironmaking, with high technology readiness. The underlying mechanisms governing this process are characterized by a complex interaction of several chemical (phase transformations), physical (transport), and mechanical (stresses) phenomena. Their interplay leads to rich microstructures, characterized by a hierarchy of defects ranging across several orders of magnitude in length, including vacancies, dislocations, internal interfaces, and free surfaces in the form of cracks and pores. These defects can all act as reaction, nucleation, and diffusion sites, shaping the overall reduction kinetics. A clear understanding of the roles and interactions of these dynamically-evolving nano-/microstructure features is missing. Gaining better insights in these effects could enable improved access to the microstructure-based design of more efficient HyDR methods, with potentially high impact on the urgently needed decarbonization in the steel industry.

*Keywords*: Hydrogen metallurgy, Direct reduction, Iron oxides, Microstructure, Multiscale

1. **Introduction**

Steel is the most important engineering metallic material, used across countless applications in transport, infrastructure, and energy conversion. Its production of >1.8 billion tons per year, with the largest fraction reduced from iron oxides using carbon, leads to huge anthropogenic $CO_2$ emissions, accounting for ~30% of all industrial $CO_2$ emissions (~7% of the total $CO_2$ emissions) [1, 2]. These numbers qualify steelmaking as the largest single cause of global warming. This scenario poses an urgent decarbonization challenge [3]. Numerous efforts have been made to mitigate $CO_2$ emissions, by improving efficiency and storing $CO_2$ underground [4, 5]. However, drastic $CO_2$ reduction cannot be realized by existing technology, but more disruptive approaches must be studied and upscaled [2, 6-8]. Along the steel production chain, the reduction of iron ore in blast furnaces using carbon (via coke and coal) causes alone as much as ~80-90% of the $CO_2$ emissions [8, 9]. Thus, fossil-fuel-free ironmaking has the biggest leverage for the reduction of $CO_2$ emissions. Hydrogen-based direct reduction (HyDR) of iron oxides is one of the most promising solutions in this context. $CO_2$ emissions below 0.1 tons per ton of steel are estimated if hydrogen is produced using renewable energy, in contrast to ~1.9 tons $CO_2$ per ton of steel emitted from the blast furnace and basic oxygen furnace route [10]. In HyDR, pelletized solid iron ores (hematite, magnetite) are reduced by hydrogen gas. The reduction steps from hematite towards iron strongly depend on temperature [2, 11]. When the reduction temperature is >570 °C, the reactions evolve in the sequence hematite ($Fe_2O_3$) → magnetite ($Fe_3O_4$) → wüstite ($Fe_{(1-x)}O$, where x indicates Fe deficiency) → iron (α-Fe or γ-Fe). For reduction <570 °C, wüstite is thermodynamically unstable and the reduction proceeds via hematite → magnetite → α-iron [2]. As such, HyDR is a multistep solid-gas reaction where the solid undergoes several phase transformations.

Many studies addressed the influence of processing parameters (*e.g.*, temperature [12, 13], gas composition [14, 15], and pressure [16, 17]) on the kinetics of the gaseous reduction of iron oxide with CO, $H_2$, $CH_4$, and their mixtures. In general, HyDR proceeds two to three times faster than CO-based direct reduction. This is attributed to the physical properties of hydrogen, *i.e.*, its small molecule size, low viscosity and high mobility when diffusing as molecule through pores or as dissociated atom through solids [2]. The HyDR process is characterized by a hierarchy of phenomena that can influence the reaction at different length and time scales. They range from transport and reaction kinetics in a shaft reactor at macroscopic scale down to chemical reactions at interfaces at atomic scale and catalysis, dissociation and charge transfer at electronic scale. Reaction kinetics is also affected by micro-to-atomic-scale features of the different oxides and the adjacent iron layers, including crystal defects, porosity, mechanics, and local composition [2, 6, 18]. Although these dynamically evolving nano-/microstructure features can alter the kinetics by orders of magnitude, a complete picture of their roles and interactions is still missing.

This viewpoint paper presents an overview of the hierarchical nature of HyDR, revealing its complexity in terms of the multiple chemical reactions and physical phenomena at different length and time scales. Particularly, the important role of the nano-/microstructures in HyDR is highlighted, aiming to identify research tasks at the nexus between physical metallurgy and process metallurgy. We also discuss approaches for multiscale and *in-operando* characterization, integrated computational materials engineering approaches, and machine learning in this field.

## 2. Hierarchy levels in hydrogen-based direct reduction (HyDR)

The high number of process variables in HyDR, including temperature, gas composition and pressure, chemistry and size of ores, friction, sticking and ore porosity creates a system with many degrees of freedom. Quantitative prediction of the kinetics and metallization is so far not possible, because of this complexity. This is shown for the shaft furnace process in *Fig. 1*. Lack of understanding of the hierarchy of the individual processes, transport and reaction steps hinders modeling HyDR with regards to efficiency, kinetics and metallic yield.

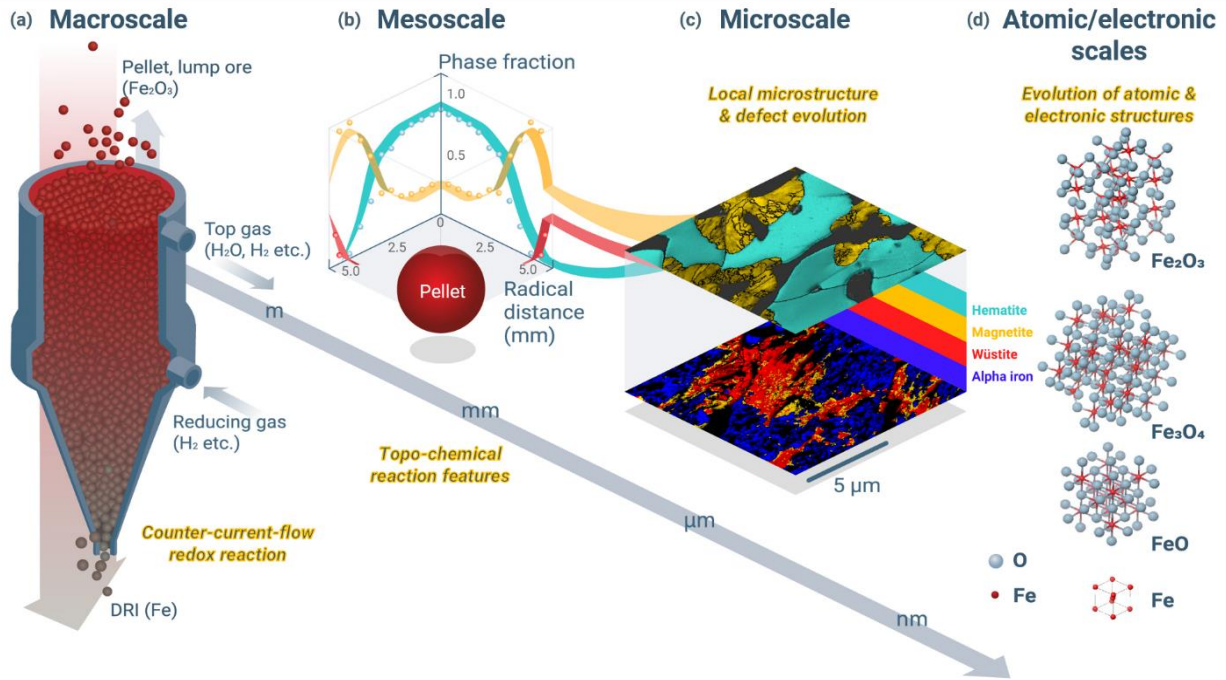

***Fig. 1*** *Hierarchical nature of hydrogen-based direct reduction of iron oxides, exemplarily shown for the shaft furnace process. (a) At the macroscopic scale, the reduction in a shaft furnace proceeds via a counter-current-flow redox reaction and the overall reduction degree is a function of furnace height and radius. (b) At the mesoscopic scale, the iron oxide pellet is reduced depending on its topo-chemical features, i.e., phase transformations and mass transport occur in gradients as a function of distance from the pellet surface towards its center (along radius). At the (c) microscopic and (d) nanoscopic/electronic scales, the reduction takes place at the reacting interfaces, associated with the multiple physical and chemical phenomena, e.g., multistep solid-gas reactions, phase transformation, solid-state diffusion, generation of microscopic defects and porosity, where the microstructure evolves in a spatiotemporal manner, as a function of reduction time and position in the pellet. (DRI stands for direct reduced iron.)*

## 2.1. Macroscopic reduction behavior

Shaft furnaces for HyDR are operated as counter-current flow reactors (***Fig. 1a***). The ores, *i.e.*, burden, in the form of pellets or lump, are charged at the top and move downwards, while the reducing gas percolates upward through the burden [8]. The processing parameters (*e.g.*, temperature, gas composition, and burden topology) and the reduction kinetics and yield are therefore functions of furnace height and radius. Several mean-field simulations [8] and experiments [19] have been conducted to investigate the effect of these parameters on the reduction kinetics at shaft furnace scale. The predictive capabilities of such macroscopic descriptions remain, however, limited due to the complex hierarchical nature of the underlying reaction steps. This includes such important questions as optimal pellet size, friction, abrasion and sticking.

## 2.2. Mesoscopic reduction behavior

At the mesoscopic scale, the reduction is determined by the local thermo-chemical boundary conditions and redox kinetics of a single oxide piece or pellet. A pellet is an agglomerate of dense particles of iron oxides, which are surrounded by pores and other oxides from the gangue and pellet binder. A single particle comprises irregularly shaped crystallites of iron oxides. The solid-gas reaction proceeds in a topo-chemical fashion at the pellet scale, *i.e.*, phase transformations and mass transport occur in gradients as a function of the distance from the pellet surface towards its center (***Fig. 1b***) [20].

## 2.3. Micro-to-atomic-scale reduction behavior

The mesoscopic reduction is determined by microscopic and atomistic mechanisms, with multiple interacting chemical, physical, and mechanical phenomena. These include multistep solid-gas reactions, mass and electron transport through heterogeneous media, multiple phase transformations, and intense mechanical interaction among the phases. This leads to the evolution of a complex defect cosmos, including vacancies, dislocations, interfaces, and free surfaces (cracks, pores), all with different transport features. These defects act as reaction, nucleation, and diffusion sites, shaping the overall reduction kinetics. Understanding these mechanisms is thus necessary for modeling and designing efficient reactors, that can operated with variable feedstock. Also, the roles of gangue chemistry and catalytic effects are unexplored.

## 3. Hydrogen-based direct reduction of iron oxides at microscopic and nanoscopic scales

## 3.1. Physics and chemistry fundamentals

The microscopic reaction steps during HyDR are similar to other oxide reductions with hydrogen [21]. At high temperatures, $H_2$ is dissociatively absorbed into vacant interstitial sites at the oxide surface, then $OH^-$ is formed until the second H is bonded and the reaction is completed. Next the water is either adsorbed to the surface or evaporates into the gas phase. The reaction sequence can be summarized as:

$$H_2 + 2Va^* \rightarrow 2H^* \tag{1}$$

$$H^* + O^{2-} \rightarrow OH^- + e^- \tag{2}$$

$$OH^- + H^* \rightarrow H_2O + e^- \tag{3}$$

where Va* indicates a vacant interstitial site. This sequence depends on the surface, bulk, and defect diffusion of $H^*$ and $O^{2-}$. The most probable site for this reduction sequence is at the solid surface. Catalysis might alter these steps, an effect that depends on gangue-related elements in the ores.

Several studies have discussed elementary processes of diffusion and reactions at internal phase boundaries [6, 22-24]. Fundamental differences exist between the oxide/oxide interfaces (*e.g.*, hematite/magnetite and magnetite/wüstite) and the oxide/iron interfaces (*e.g.*, wüstite/iron). ***Fig. 2****(a)* shows the reduction of hematite to magnetite. Oxygen is removed by hydrogen (which must first dissociate into mono-atomic form) on a free surface of an oxide surrounded by open-pore channels. An activity gradient of iron ions ($Fe^{2+}$) is built up across magnetite. Iron ions and electrons migrate to the internal reaction interfaces, where they react with hematite to produce magnetite based on the following reaction [24, 25]:

$$4Fe_2O_3 + Fe^{2+} + 2e^- \rightarrow 3Fe_3O_4 \tag{4}$$

Analogously, when the magnetite is reduced to wüstite the reaction at the magnetite/wüstite interface

can be expressed as:

$$Fe_3O_4 + Fe^{2+} + 2e^- \rightarrow 4FeO \qquad (5)$$

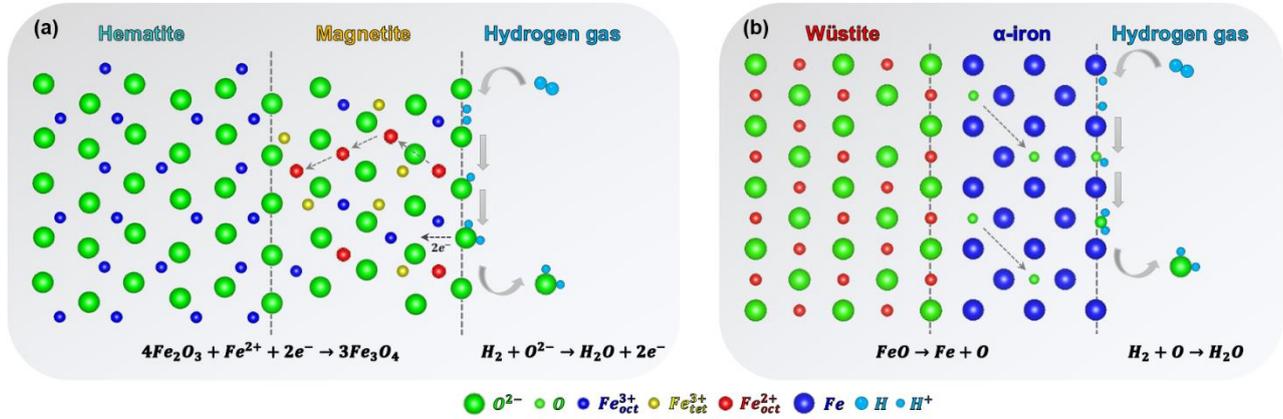

*Fig. 2* Schematic illustration of hydrogen-based direct reduction mechanisms at the atomic scale: (a) reduction of trigonal hematite to cubic magnetite [25] and (b) reduction of cubic wüstite to cubic α-iron [6].

In contrast, oxygen atoms must diffuse from the iron/wüstite interface to the iron/gas interface once a dense iron layer forms, *Fig. 2(b)* [6, 18]. Then, hydrogen reacts with oxygen at the iron/gas interface, forming water. The removal, storage, dissociation, or adsorption behavior of the water at/from the interfaces is unexplored terrain, yet, with high relevance for reduction. The driving force for outbound oxygen diffusion is the gradient of the oxygen activity from the internal iron/wüstite interface through the iron layer to the external iron/gas interface. At the internal reacting interface, wüstite dissociation into iron and oxygen prevails according to the following reaction [6]:

$$FeO \rightarrow Fe + O \qquad (6)$$

The oxygen solubility and diffusivity in iron are small, *e.g.*, at 700 °C in α-iron, 4.65×10$^{-5}$ wt.% and 2.16×10$^{-11}$ m$^2$·s$^{-1}$, respectively (ThermoCalc, TCOX10 [26]). Therefore, the oxygen diffusion through dense iron is only possible for very thin layers. However, the reaction can proceed through alternative transport pathways inside the iron, along cracks, pores, and other defects with high diffusion coefficients such as dislocations and grain boundaries.

Some of the observed crystallographic orientation relationships (OR) suggest plausible phase transformation mechanisms at the internal interphase boundaries [23, 27]. A Shoji-Nishiyama OR, *i.e.*, $(0001)_{hem}//(111)_{mag}$ and $[10\bar{1}0]_{hem}//[1\bar{1}0]_{mag}$, was observed at the hematite/magnetite interface. This OR implies a shear transformation mechanism of oxygen lattices from hematite to magnetite associated with the rearrangement of iron ions in the interstitial sites by diffusion [27]. At the magnetite/wüstite interface, a cube-on-cube OR (*i.e.*, $(1\bar{1}0)_{mag}//(1\bar{1}0)_{wüs}$ and $[110]_{mag}//[110]_{wüs}$) was observed [23]. This OR indicates that reduced iron ions (*e.g.*, on the external solid/gas interface) migrate towards ferrous vacancies on oxygen ion planes to form wüstite, while the oxygen lattice is maintained [23]. There is still a lack of microscopic and spectroscopic evidence as well as atomistic understanding of these diffusion and reaction mechanisms across the interfaces. Here, diffusion couples with isotope labeling could provide further insights [28].

## 3.2. Influence of chemical composition of the feedstock

As HyDR is a solid state reaction, the gangue elements in the burden remain inside the reduced iron, in part as unreduced oxides. To produce not only green but also clean steel, the gangue elements and their oxides must be removed during subsequent steelmaking, requiring energy and costs. Thus, high-quality ores with high iron (>67 wt.%) and low tramp element content are desirable as feedstock for HyDR [29]. However, gangue elements such as silicon, calcium, magnesium and aluminum occur in most ores and can have effects on the reduction. Silicon and manganese combine with iron oxides and form stable compounds. An example are iron-manganese silicates which are hard to reduce, lowering metallic yield [30]. Difference in thermal expansion between iron oxides and gangue oxides (*e.g.*, MnO, MgO, and CaO) results in swelling and cracking [31-33]. Small controlled quantities of gangue elements could thus even be beneficial, creating new percolation paths for mass transport of reactants and products, accelerating reduction kinetics, provided that the pellet strength is too low. Very high gangue element and inert oxide content can reduce pellet strength, resulting in fine debris which reduces permeability and causes sticking. Segregation of gangue elements at the reaction front was found to affect the local reaction kinetics [6, 34]. The gangue elements in solid solution inside the wüstite can affect the activity of iron ions and their diffusivity. For example, segregation of $Mn^{2+}$ (and $Mg^{2+}$) ions at the progressing reaction interface can decrease the activity of $Fe^{2+}$ to a value low enough to suppress formation of metallic iron [34]. In contrast, Ca segregation was found to facilitate the formation of porous iron and enhance the reduction kinetics [32, 33]. The mechanisms behind these effects of gangue elements on HyDR remain elusive. This is due to their variety and different concentrations in ores and a lack of highly resolving and *in-operando* measurements. This opens up opportunities for research about gangue elements, particularly when targeting low-grade iron ores, regarding their role on microstructure, chemical partitioning, phase formation, interface energies and catalysis.

### 3.3. Influence of crystal defects, porosity, and micromechanics

Defects in the solid feedstock can accelerate reactant transport [11, 35-39], ***Fig.* 3**. The acquired defects evolving during HyDR differ from those inherited from pelletizing. The latter determine the starting conditions, while the acquired defects gradually evolve and alter the local reduction conditions. The microstructure changes are due to (1) the mass/volume loss when oxygen is removed; (2) the non-volume conserving and non-commensurate nature of the phase transformations; (3) the local accumulation of gas (*e.g.*, water) and associated volume expansion and crack opening; (4) the cavitation and delamination at hydrogen-weakened metal/oxide interfaces; and (5) thermal spalling caused by impurities [6, 20, 40, 41]. These features depend on the boundary conditions during reduction. It was observed that porosity increases during the reduction of magnetite and wüstite with hydrogen at higher temperature and hydrogen partial pressure. The pore size increases with higher temperature and lower hydrogen partial pressure.

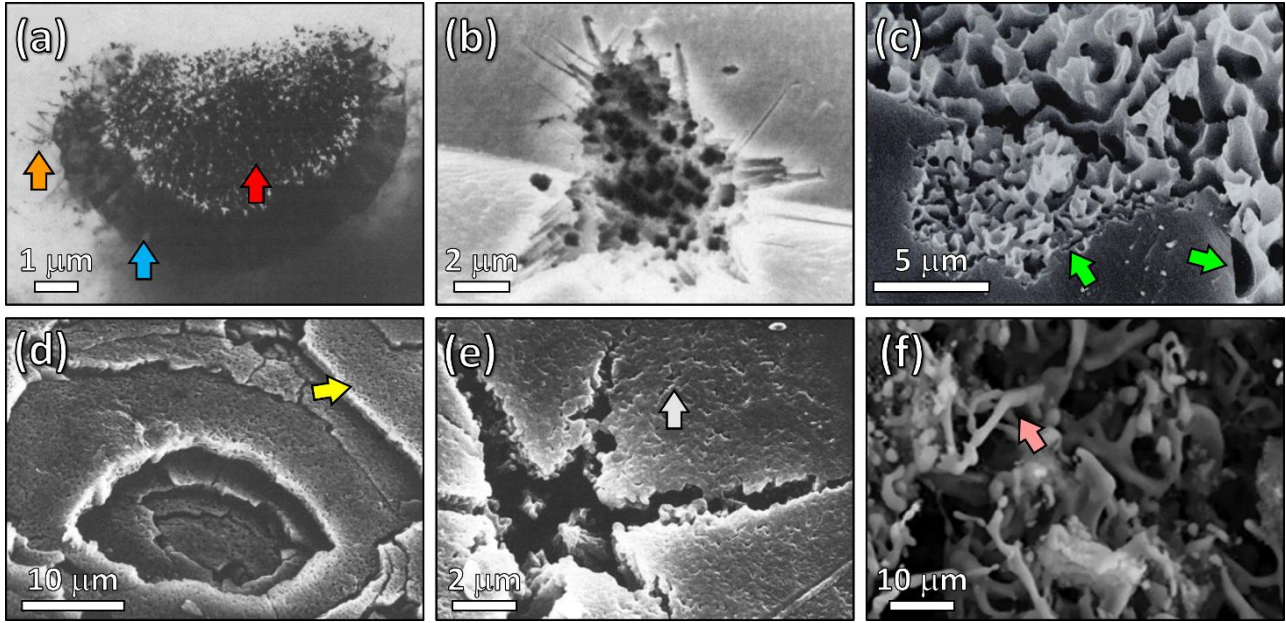

*Fig. 3* Examples of defects formed during the gaseous reduction of iron ores: (a) Reaction zones after partial reduction of hematite with pure $H_2$ at 387 °C. The black band is magnetite (blue arrow) which separates the inner structure (red arrow) composed of a mixture of porous magnetite and iron from unreacted hematite (orange arrow). The reaction interface between magnetite and hematite seems to contain dislocations [11]. (b) Wüstite partially reduced at 900 °C using a gas mixture of 70% $H_2$ and 30% $H_2O$. A large array of pits on the surface of wüstite is observed before iron nucleation [35]. (c) Partially reduced wüstite at 1000 °C using a gas mixture of 80% $H_2$ and 20% $H_2O$. The interface between iron and wüstite grows irregularly and is accompanied of pores with different sizes (green arrows) [36]. (d) Details of the surface of iron nuclei formed during reduction of magnetite at 750 °C using a hydrogen partial pressure of 26.7 kPa. Iron nucleation is accompanied by cracks that form concentrically around nucleation sites and pores (yellow arrow)[37]. (e) Details of the surface of iron nuclei formed during reduction of magnetite at 500 °C using a hydrogen partial pressure of 13.3 kPa. Iron nucleation is accompanied of micro-sized fissures with a star-like morphology and micropores (grey arrow)[37]. (f) Hematite sample fully reduced with $H_2$ up to 1000 °C at a heating rate of 20 °C/min. Iron grows in the form of whiskers (pink arrow) and pores are formed [38].

Crystal defects affect the transport of iron ions from the external solid/gas interface to the advancing solid/solid interface and of oxygen from the wüstite/iron interface to the iron/gas interface in the final reduction stage [22, 42]. The grain refinement during reduction of hematite to magnetite increases the reaction rate owing to the larger interface density [22, 42]. They are formed because of the non-plane-matching nucleation and growth of the magnetite and the high multiplicity of the plane-matching ORs between the phases. Besides the Shoji-Nishiyama OR [42, 43], however, many other ORs are not understood. Hence, crystallographic interface studies would help to understand the nucleation, growth and grain refinement mechanisms (see Section 3.1). Furthermore, geometrically necessary dislocations were observed at the wüstite/iron interfaces [6]. These can possibly act as fast diffusion channels for hydrogen (inbound) and oxygen (outbound).

Porosity is a critical factor when gaseous diffusion and local storage/transport of water are rate-limiting steps in HyDR [44, 45]. The porosity and its tortuosity affect gas percolation in and out of the internal free surface regions (*i.e.*, inbound hydrogen and outbound water transport). Internal free surfaces are also likely to promote nucleation of the phases, affecting reduction kinetics. The formation of open pores and cracks during the last reduction stage from wüstite to iron enhances solid state diffusion of oxygen [20] and gaseous diffusion of hydrogen and water. Water trapped inside isolated pores lowers the hydrogen partial pressure locally, retarding iron nucleation [6] or even causing local re-oxidation. This means that defects alter the local boundary conditions (*e.g.*, oxygen and hydrogen chemical potentials, topology, transport and temperature), affecting local thermodynamics and kinetics of the reduction.

The quantitative correlation between the defects and the reaction kinetics remains unclear, owing to the complexity of the microstructure and its evolution, Yet, this information is critical to assess the contribution of individual defects to kinetics. Therefore, the microstructure evolution during reduction should be studied and correlated to kinetic changes, with a focus on cracks and pores, possibly at near-atomic dimensions and real time. Quantification of pore size, distribution, morphology, and connectivity is needed to relate these features to transport, local reaction conditions and kinetics.

### 3.4. Advances in characterization and simulation techniques for direct reduction

The next level of insights into HyDR requires multiscale experiments and simulations (**Fig. 4**) that consider the phenomena discussed above and their interactions. Characterization with high spatial and time resolution are needed to capture the fine chemical and structural features down to atomic scale and real time. High-resolution electron microscopy is needed for probing the structure, chemistry and bonding states. Environmental microscopy can be used for *in-operando* characterization of microstructure formation. These methods could provide insights into nucleation and growth of new phases, types and positions of reactants and intermediate products, as well as the evolution of defect patterns [11]. Synchrotron facilities provide hard X-ray based techniques, such as diffraction, tomography, microscopy, and spectroscopy, to study crystal structures, connectivity of pores, defect density, and elemental distribution [46, 47]. They allow probing bulk samples with good statistics and under *in-operando* conditions [48]. Atom probe tomography (APT) allows for three-dimensional compositional mapping with sub-nanometer spatial resolution [49] and high elemental sensitivity (atomic-parts-per-million range) across all atomic masses, including hydrogen and oxygen, albeit with certain precautions [50, 51]. The development of dedicated reactors opens up opportunities for quasi-*in-situ* analysis at near atomic-scale [52, 53]. This approach allows capturing the compositional and certain crystallographic aspects of the transient states, of reaction mechanisms, segregation to interfaces and elemental partitioning. In order to reveal mechanisms that matter for large reactors, boundary conditions should be chosen in appropriate parameter ranges, to match scenarios encountered in real operation.

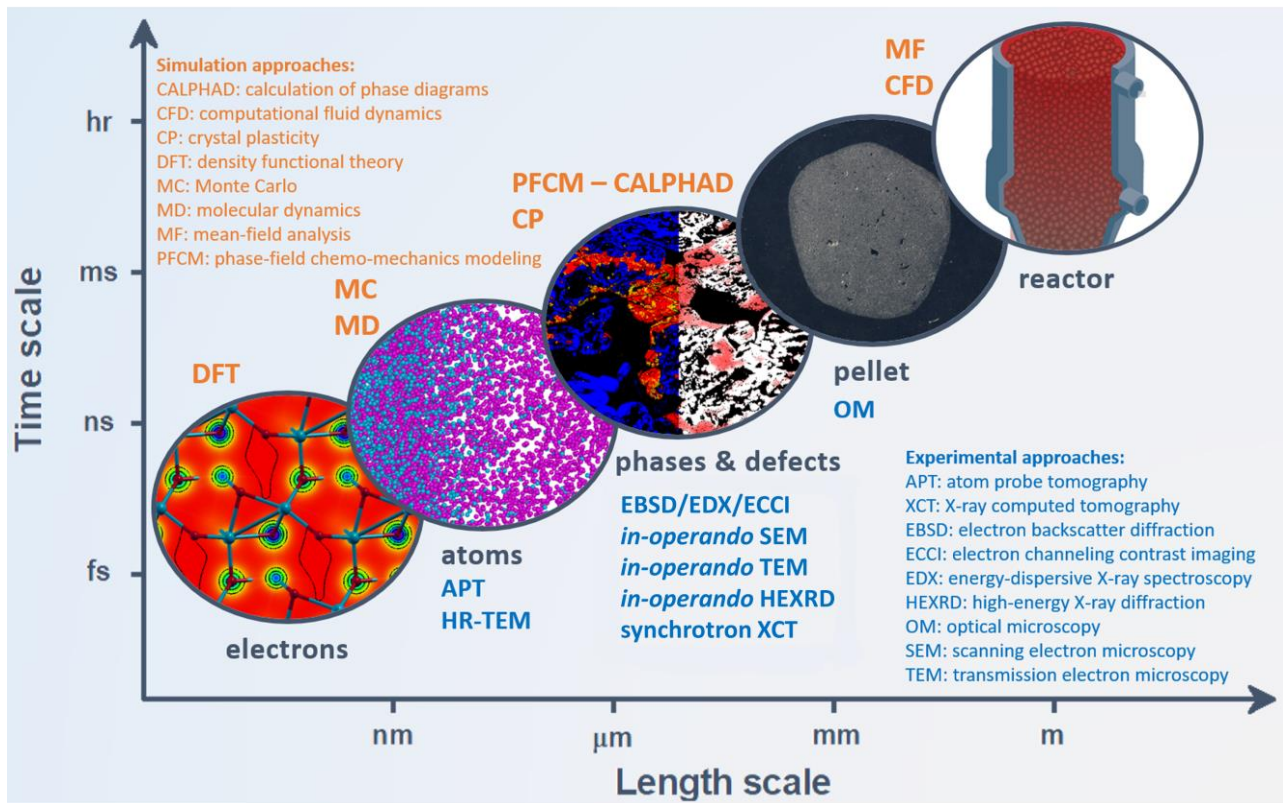

***Fig. 4*** *Multiscale characterization and simulation approaches towards better understanding of hydrogen-based direct reduction of iron oxides.*

Physics-based modeling can provide fundamental reaction details, help to interpret experimental findings, inform mean-field models, and guide the design of key experiments for HyDR. For example, HyDR of oxide to iron induces a volume change of about 42% [54], leading to residual stresses, inelastic deformation, cracks and delamination. These features affect the reactions and kinetics. When aiming to consider such multi-physics conditions, a possible modeling approach could be finite deformation phase-field chemo-mechanics (PFCM) [55]. For simulating HyDR, the method could be extended to include chemical reactions and catalysis effects [56]. Thermodynamic data obtained from density functional theory (DFT) and quantum kinetic Monte Carlo (kMC) simulations and Gibbs energy data from CALPHAD databases could serve to calibrate PFCM models for HyDR. These models would capture microscopic reduction processes and phase transformations under different boundary conditions, facilitating fast phase space and mechanism screening, for understanding the interplay of chemistry, phase transformations, transport, and mechanics. These findings could be used for improved reactor design and identification of optimal reduction conditions.

For understanding reaction mechanisms as proposed in **Fig. 2**, atomic-scale simulations are required where experiments have insufficient resolution. DFT simulations can help to investigate the surface reduction mechanisms, transition states, catalytic effects, molecular dissociation and recombination, activation energies, and charge transport [57]. *Ab initio* molecular dynamics (AIMD) can help to study the atomic structures during reduction at various temperatures and the mechanisms of oxygen removal [58]. DFT and AIMD simulations are computationally expensive and limited to system sizes of a few hundreds of atoms. Reactive force field molecular dynamics (ReaxFF-MD) [59] allows for continuous bond formation/breaking. It is suited to model chemical reactions with more a few million atoms [60]. With carefully fitted potentials, ReaxFF-MD could provide insights into the nano-chemistry of HyDR, including the evolution of defects and species diffusion. Combinations of DFT, MD, and machine learning (ML) might be promising to reach the accuracy of quantum mechanical computations at reduced computational costs [61].

ML might also help to accelerate or replace standard solvers for the partial differential equations of such models [62]. It could also automate the identification of constitutive parameters for such complex systems, from the analysis of kinetic experimental data and microscopy images [63]. This might enhance multiscale and multiphysics simulations by orders of magnitude and help solving such complex optimization and inverse problems as in HyDR.

## 4. Summary, outlook and research opportunities

Hydrogen-based direct reduction is a promising approach for green iron making, mitigating $CO_2$ emissions in this sector. The underlying processes are characterized by the nonlinear interplay of several chemical, physical, and mechanical phenomena that extend across several length and time scales, establishing a hierarchical nature of this redox process. Advanced characterization and simulation tools allow to map and understand these phenomena at relevant scales and in real time. This helps to tune ore and pellet properties, choose appropriate feedstock, design reactors, and optimize process conditions. These points are important to overcome the kinetic limits of this process, enabling more efficient carbon-free reduction, closer to the thermodynamic and kinetic limits.

Main research opportunities can be grouped into advanced characterization; theory, simulation, and machine learning; roles of oxides and their chemical composition; use of pellets vs. fine or lump oxides with respect to different reactors; reduction gases and their mixtures; microstructure; influence of microstructure gradients and feedstock size effects; as well as phase transformation under non-equilibrium conditions.

- Advanced *in-situ* and *operando* characterization of the reduction at typical reactor operation temperatures between 500 and 1000 °C and with different hydrogen partial pressures from the atomic to mesoscopic scales forms the backbone approach in this field, to reveal the elementary mechanisms in a quantitative fashion. Suited methods and approaches range from quasi-*in-situ* atom probe tomography, environmental microscopy, to synchrotron- and neutron-based probing methods.
- Simulation, theory, and machine learning assume a parallel role in interrogating the

experimental *in-situ* observations and putting theoretical questions forward. Particularly, multi-scale simulations will play an essential role in bridging the length scales from the atomic scale where the elementary redox reaction occurs up to the reactor level where macroscopic gas transport and gradients in the local boundary conditions become relevant. The atomistic simulations seem particularly challenging, as some of the elementary chemical reaction and transport steps, influence of impurities, catalysis effects, and the nucleation during the multiple associated phase transformations are rare event phenomena when measured by atomic relaxation scales, rendering the atomistic simulation scale in itself a massive multiscale challenge, particularly regarding time. A benefit of such efforts is that these elementary steps of this redox reaction play a huge role in several science fields beyond steelmaking, such as geophysics, catalysis, electrochemistry, or corrosion, making them a worthwhile goal for basic research. Machine learning methods can greatly enhance and support all of these simulation challenges, for the development of better interatomic potentials or as faster and more efficient solvers when it comes to the integration of the underlying partial differential equations that serve to formulate coupled mean-field constitutive transport, reaction, phase transformation, and mechanical models.

- An essential task is the study of the metallization and reduction rates achievable for different oxide and pellet types under consideration of their chemical composition, magnetic properties, and gangue elements, with respect to the resulting slag formation when the sponge iron is charged into electric arc furnaces and regarding possible contamination effects on the final steel.

- A related task is the investigation of the metallic yield when using pellets versus fine or lump oxides, under consideration of the different possible reactor design types from static direct-reduction to fluidized-bed reactors. This includes also the effects of the size distribution, friction, abrasion and sticking behavior of the pellets or respectively ores for the individual reactor concepts.

- Along with these solid feedstock variants, the investigation of different types of H-carrying reduction gases and their mixtures is an important task, including the efficiency in the use of hydrogen. In this context, two topics seem to be of particular relevance. The first one is the use of alternative hydrogen-containing reduction gases such as ammonia or liquid organic hydrogen carriers. Such hydrogen carriers can not only release the required hydrogen but also offer more efficient means for global storage and transport of intermittent renewable energy. This is an important aspect in light of the huge annual steel production, which requires likewise huge quantities of energy and reductants, as well as their transcontinental transport. The second aspect is the efficiency in the exploitation of the hydrogen in a real-world reduction process, as this reductant is an economical key factor. Thus, any reactor and process should be designed in a way to exploit it as near-stoichiometric as possible.

- The microstructure plays a particularly important role in this field for three main reasons. First, phase transformations during reduction lead to a very complex evolution of the associated microstructures. Second, the different microstructure ingredients, particularly the interfaces and the dislocations, are important pathways for accelerated transport of the reaction species. Third, one important kinetic barrier is the nucleation during the phase transformation sequence, where particularly the iron formation is a sluggish process. In this context, the lattice defects assume a critical role as they offer sites for heterogeneous nucleation, lowering the required nucleation barriers.

- Influence of microstructure gradients and oxide feedstock size effects play also an eminent role. Recent work has shown that size effects and gradients in the evolving microstructure of the feedstock material are important for metallization and reduction kinetics. Originally, this effect was assumed to be of minor relevance owing to the fast diffusion of hydrogen. However, the transport of oxygen and the removal of the water from the reaction front were found to be highly microstructure- and size-dependent, thus causing kinetic bottlenecks. Consequently, a detailed investigation of the ideal pellet and/or oxide sizes, the underlying microstructures, and their gradients through the bulk feedstock are important research topics for optimizing direct reduction kinetics and metallization.

- New insights can also be expected from the investigation of the atomistic mechanisms behind the phase transformations that proceed under chemical non-equilibrium conditions. Through the gradual removal of oxygen, the system undergoes an unusually large number of sequential phase transformations. The underlying nucleation and growth processes under such chemically graded conditions are not yet well understood. Like some of the other phenomena mentioned, these investigations can also shed light on related research fields, including geology, corrosion, electrochemistry, and catalysis.

**Acknowledgments**

Y. Ma acknowledges financial support through Walter Benjamin Programme of the Deutsche Forschungsgemeinschaft (Project No. 468209039). I.R. Souza Filho acknowledges financial support through CAPES (Coordenação de Aperfeiçoamento de Pessoal de Nível Superior - Brazil) & Alexander von Humboldt Foundation (Project No. 88881.512949/2020-01). D. Xie acknowledges the financial support from the Alexander von Humboldt Foundation. B. Gault acknowledges financial support from the ERC-CoG-SHINE-771602.

**References**

[1] D. Raabe, C.C. Tasan, E.A. Olivetti, Nature 575(7781) (2019) 64-74.
[2] D. Spreitzer, J. Schenk, Steel Res. Int. 90(10) (2019) 1900108.
[3] T. Ariyama, K. Takahashi, Y. Kawashiri, T. Nouchi, J. Sustain. Met. 5(3) (2019) 276-294.
[4] J. van der Stel, G. Louwerse, D. Sert, A. Hirsch, N. Eklund, M. Pettersson, Ironmak. Steelmak. 40(7) (2013) 483-489.
[5] S. Pauliuk, R.L. Milford, D.B. Müller, J.M. Allwood, Environ. Sci. Technol 47(7) (2013) 3448-3454.
[6] S.-H. Kim, X. Zhang, Y. Ma, I.R. Souza Filho, K. Schweinar, K. Angenendt, D. Vogel, L.T. Stephenson, A.A. El-Zoka, J.R. Mianroodi, M. Rohwerder, B. Gault, D. Raabe, Acta Mater. 212 (2021) 116933.
[7] I.R. Souza Filho, Y. Ma, M. Kulse, D. Ponge, B. Gault, H. Springer, D. Raabe, Acta Mater. 213 (2021) 116971.
[8] F. Patisson, O. Mirgaux, Metals 10(7) (2020) 922.
[9] M. Kundak, L. Lazić, J. Črnko, Metalurgija 48(3) (2009) 193-197.
[10] V. Vogl, M. Åhman, L.J. Nilsson, J. Clean. Prod. 203 (2018) 736-745.
[11] M.-F. Rau, D. Rieck, J.W. Evans, Metall. Trans. B 18(1) (1987) 257-278.
[12] M. Bahgat, M.H. Khedr, Mater. Sci. Eng. B 138(3) (2007) 251-258.
[13] A.A. El-Geassy, M.I. Nasr, Trans. Iron Steel Inst. Jpn. 28(8) (1988) 650-658.
[14] A.A. El-Geassy, V. Rajakumar, Trans. Iron Steel Inst. Jpn. 25(6) (1985) 449-458.
[15] A. Bonalde, A. Henriquez, M. Manrique, ISIJ Int. 45(9) (2005) 1255-1260.
[16] A. Habermann, F. Winter, H. Hofbauer, J. Zirngast, J.L. Schenk, ISIJ Int. 40(10) (2000) 935-942.
[17] K. Sato, Y. Ueda, Y. Nishikawa, T. Goto, Trans. Iron Steel Inst. Jpn. 26(8) (1986) 697-703.
[18] R.J. Fruehan, Y. Li, L. Brabie, E.-J. Kim, Scand. J. Metall. 34(3) (2005) 205-212.
[19] M. Pei, M. Petäjäniemi, A. Regnell, O. Wijk, Metals 10(7) (2020) 972.
[20] G. Bitsianes, T.L. Joseph, JOM 7(5) (1955) 639-645.
[21] Y. Bai, M. Mavrikakis, J. Phys. Chem. B 122(2) (2018) 432-443.
[22] J.O. Edström, J. Iron Steel Inst. 175 (1953) 289-304.
[23] X. Guo, Y. Sasaki, Y. Kashiwaya, K. Ishii, Metall. Mater. Trans. B 35(3) (2004) 517-522.
[24] A. Zare Ghadi, M.S. Valipour, S.M. Vahedi, H.Y. Sohn, Steel Res. Int. 91(1) (2019) 1900270.
[25] C. Feilmayr, A. Thurnhofer, F. Winter, H. Mali, J. Schenk, ISIJ Int. 44(7) (2004) 1125-1133.
[26] J.O. Andersson, T. Helander, L. Höglund, P. Shi, B. Sundman, Calphad 26(2) (2002) 273-312.
[27] Y. Watanabe, S. Takemura, Y. Kashiwaya, K. Ishii, J. Phys. D: Appl. Phys 29(1) (1996) 8-13.
[28] T.C. Kaspar, S.D. Taylor, K.H. Yano, T.G. Lach, Y.D. Zhou, Z.H. Zhu, A.A. Kohnert, E.K. Still, P. Hosemann, S.R. Spurgeon, D.K. Schreiber, Adv. Mater. Interfaces 8(9) (2021) 2001768.
[29] L. Lu, J. Pan, D. Zhu, in: L. Lu (Ed.), Iron Ore, Woodhead Publishing2015, pp. 475-504.
[30] A. El-Geassy, M. Nasr, A.A. Omar, E.A. Mousa, ISIJ Int. 48 (2008) 1359-1367.
[31] M. Moukassi, M. Gougeon, P. Steinmetz, B. Dupre, C. Gleitzer, Metall. Trans. B 15(2) (1984) 383-391.
[32] Y. Iguchi, K. Goto, S. Hayashi, Metall. Mater. Trans. B 25(3) (1994) 405-413.
[33] Y. Iguchi, Y.-I. Ueda, S. Hayashi, Metall. Mater. Trans. B 25(5) (1994) 741-748.

[34] D.S. Chen, B. Song, L.N. Wang, T. Qi, Y. Wang, W.J. Wang, Miner. Eng. 24(8) (2011) 864-869.
[35] D.H.S. John, P.C. Hayes, Metall. Trans. B 13(1) (1982) 117-124.
[36] S.P. Matthew, T.R. Cho, P.C. Hayes, Metall. Trans. B 21(4) (1990) 733-741.
[37] S.P. Matthew, P.C. Hayes, Metall. Trans. B 21(1) (1990) 153-172.
[38] A. Hammam, Y. Li, H. Nie, L. Zan, W. Ding, Y. Ge, M. Li, M. Omran, Y. Yu, Mining Metall. Explor. 38(1) (2021) 81-93.
[39] A. Ranzani Da Costa, Reduction of iron ore by H2: kinetics, sticking, and modeling La réduction du minerai de fer par l'hydrogène : étude cinétique, phénomène de collage et modélisation, Institut National Polytechnique de Lorraine (INPL), 2011.
[40] D.G. Xie, Z.J. Wang, J. Sun, J. Li, E. Ma, Z.W. Shan, Nat. Mater. 14(9) (2015) 899-903.
[41] M. Li, D.G. Xie, E. Ma, J. Li, X.X. Zhang, Z.W. Shan, Nat. Commun. 8 (2017) 14564.
[42] R.O. Keeling, D.A. Wick, Science 141(3586) (1963) 1175-1176.
[43] L.A. Bursill, R.L. Withers, J. Appl. Crystallogr. 12(3) (1979) 287-294.
[44] R.G. Olsson, W.M. McKewan, Metall. Trans. 1(6) (1970) 1507-1512.
[45] E.T. Turkdogan, R.G. Olsson, J.V. Vinters, Metall. Mater. Trans. B 2(11) (1971) 3189-3196.
[46] J.C. Tseng, D. Gu, C. Pistidda, C. Horstmann, M. Dornheim, J. Ternieden, C. Weidenthaler, Chemcatchem 10(19) (2018) 4465-4472.
[47] K. Bugelnig, P. Barriobero-Vila, G. Requena, Pract. Metallogr. 55(8) (2018) 556-568.
[48] K.-D. Liss, A. Bartels, A. Schreyer, H. Clemens, Textures Microstruct. 35(3-4) (2003) 219-252.
[49] B. Gault, A. Chiaramonti, O. Cojocaru-Mirédin, P. Stender, R. Dubosq, C. Freysoldt, S.K. Makineni, T. Li, M. Moody, J.M. Cairney, Nat. Rev. Methods Primers 1(1) (2021) 51.
[50] A.J. Breen, L.T. Stephenson, B. Sun, Y. Li, O. Kasian, D. Raabe, M. Herbig, B. Gault, Acta Mater. 188 (2020) 108-120.
[51] M. Bachhav, F. Danoix, B. Hannoyer, J.M. Bassat, R. Danoix, Int. J. Mass Spectrom. 335 (2013) 57-60.
[52] H. Khanchandani, A.A. El-Zoka, S.-H. Kim, U. Tezins, D. Vogel, A. Sturm, D. Raabe, B. Gault, L. Stephenson, Laser-equipped gas reaction chamber for probing environmentally sensitive materials at near atomic scale, 2021, https://arxiv.org/abs/2107.11987.
[53] P.A.J. Bagot, T. Visart de Bocarmé, A. Cerezo, G.D.W. Smith, Surf. Sci. 600(15) (2006) 3028-3035.
[54] W. Mao, W.G. Sloof, Metall. Mater. Trans. B 48(5) (2017) 2707-2716.
[55] B. Svendsen, P. Shanthraj, D. Raabe, J. Mech. Phys. Solids 112 (2018) 619-636.
[56] Y. Bai, J.R. Mianroodi, Y. Ma, A.K. da Silva, B. Svendsen, D. Raabe, Chemo-Mechanical Phase-Field Modeling of Iron Oxide Reduction with Hydrogen, 2021, https://arxiv.org/abs/2111.11538.
[57] H. Zhong, L. Wen, J. Li, J. Xu, M. Hu, Z. Yang, Powder Technol. 303 (2016) 100-108.
[58] R. Rousseau, V.-A. Glezakou, A. Selloni, Nat. Rev. Mater. 5(6) (2020) 460-475.
[59] A.C.T. van Duin, S. Dasgupta, F. Lorant, W.A. Goddard, J. Phys. Chem. A 105(41) (2001) 9396-9409.
[60] T.P. Senftle, S. Hong, M.M. Islam, S.B. Kylasa, Y. Zheng, Y.K. Shin, C. Junkermeier, R. Engel-Herbert, M.J. Janik, H.M. Aktulga, T. Verstraelen, A. Grama, A.C.T. van Duin, Npj Comput. Mater. 2(1) (2016) 15011.
[61] P. Friederich, F. Häse, J. Proppe, A. Aspuru-Guzik, Nat. Mater. 20(6) (2021) 750-761.
[62] J.R. Mianroodi, N. H. Siboni, D. Raabe, Npj Comput. Mater. 7(1) (2021) 99.
[63] J.R. Mianroodi, S. Rezaei, N.H. Siboni, B.-X. Xu, D. Raabe, Lossless Multi-Scale

Constitutive Elastic Relations with Artificial Intelligence, 2021, http://arxiv.org/abs/2108.02837.